\documentclass[%
 reprint,
superscriptaddress, 
 amsmath,amssymb,
 aps,
 prc,
 twocolumn,
prb, 
floatfix,
]{revtex4-2}

\usepackage[utf8]{inputenc}
\usepackage{color}
\usepackage{graphicx}
\usepackage{textcomp}
\usepackage{gensymb}
\usepackage{threeparttable}
\usepackage{booktabs}
\usepackage{subfigure}
\usepackage{multirow,booktabs}

\usepackage[colorlinks,linkcolor=blue,anchorcolor=blue,citecolor=blue,urlcolor=blue,breaklinks]{hyperref}
\usepackage{url}
\usepackage[hyphenbreaks]{breakurl}

\usepackage{tikz,xcolor}
\definecolor{lime}{HTML}{A6CE39}
\DeclareRobustCommand{\orcidicon}{
\begin{tikzpicture}
\draw[lime, fill=lime] (0,0)
circle[radius=0.12]
node[white]{{\fontfamily{qag}\selectfont \tiny \.{I}D}};
\end{tikzpicture}
\hspace{-3mm}
}

\foreach \x in {A, ..., Z}{%
\expandafter\xdef\csname orcid\x\endcsname{\noexpand\href{https://orcid.org/\csname orcidauthor\x\endcsname}{\noexpand\orcidicon}}
}

\begin{document}

\title{Observation of the $\pi^2\sigma^2$-bond linear-chain molecular structure in $^{16}$C}

\def\PKU{School of Physics and State Key Laboratory of Nuclear Physics and Technology, Peking University, Beijing 100871, China}
\def\IMCAEP{Institute of Materials, China Academy of Engineering Physics, Mianyang, 621907, China}
\def\KIT{Kitami Institute of Technology, 090-8507 Kitami, Japan}
\def\PHU{Department of Physics, Hokkaido University, 060-0810 Sapporo, Japan}
\def\SHU{School of Science, Huzhou University, Huzhou 313000, China}
\def\IMP{Institute of Modern Physics, Chinese Academy of Sciences, Lanzhou 730000, China}
\def\HEU{Fundamental Science on Nuclear Safety and Simulation Technology Laboratory, Harbin Engineering University, Harbin 150001, China}
\def\RCNP{Research Center for Nuclear Physics, Osaka University, 10-1 Mihogaoka, Ibaraki 567-0047, Japan.}

\author{J. X. Han\hspace{-1.5mm}\orcidA{}} 
\affiliation{\PKU}

\author{Y. Liu} 
\email{liuyang\_enphy@pku.edu.cn}
\affiliation{\PKU}
\affiliation{\IMCAEP}

\author{Y. L. Ye\hspace{-1.5mm}\orcidB{}} 
\email{yeyl@pku.edu.cn}
\affiliation{\PKU}

\author{J. L. Lou} 
\affiliation{\PKU}

\author{X. F. Yang\hspace{-1.5mm}\orcidC{}}
\affiliation{\PKU}

\author{T. Baba}
\affiliation{\KIT}

\author{M. Kimura}
\affiliation{\PHU}

\author{B. Yang}
\affiliation{\PKU}

\author{Z. H. Li}
\affiliation{\PKU}

\author{Q. T. Li}
\affiliation{\PKU}

\author{J. Y. Xu}
\affiliation{\PKU}

\author{Y. C. Ge}
\affiliation{\PKU}

\author{H. Hua}
\affiliation{\PKU}

\author{Z. H. Yang}
\affiliation{\RCNP}

\author{J. S. Wang}
\affiliation{\SHU}
\affiliation{\IMP}

\author{Y. Y. Yang}
\affiliation{\IMP}

\author{P. Ma}
\affiliation{\IMP}

\author{Z. Bai}
\affiliation{\IMP}

\author{Q. Hu}
\affiliation{\IMP}

\author{W. Liu}
\affiliation{\PKU}

\author{K. Ma}
\affiliation{\PKU}

\author{L. C. Tao}
\affiliation{\PKU}

\author{Y. Jiang}
\affiliation{\PKU}

\author{L. Y. Hu}
\affiliation{\HEU}

\author{H. L. Zang}
\affiliation{\PKU}

\author{J. Feng}
\affiliation{\PKU}

\author{H. Y. Wu}
\affiliation{\PKU}

\author{S. W. Bai}
\affiliation{\PKU}

\author{G. Li}
\affiliation{\PKU}

\author{H. Z. Yu}
\affiliation{\PKU}

\author{S. W. Huang}
\affiliation{\PKU}

\author{Z. Q. Chen}
\affiliation{\PKU}

\author{X. H. Sun}
\affiliation{\PKU}

\author{J. J. Li}
\affiliation{\PKU}

\author{Z. W. Tan}
\affiliation{\PKU}

\author{Z. H. Gao}
\affiliation{\IMP}

\author{F. F. Duan}
\affiliation{\IMP}

\author{J. H. Tan}
\affiliation{\HEU}

\author{S. Q. Sun}
\affiliation{\HEU}

\author{Y. S. Song}
\affiliation{\HEU}

\begin{abstract}
Measurements of the $^2$H($^{16}$C,$^{16}$C$^{*}$$\rightarrow^4$He+$^{12}$Be or $^6$He+$^{10}$Be)$^2$H inelastic excitation and cluster-decay reactions have been carried out at a beam energy of about 23.5 MeV/u. A specially designed detection system, including one multi-layer silicon-strip telescope at around zero degrees, has allowed the high-efficiency three-fold coincident detection and therefore the event-by-event determination of the energy of the unstable nucleus beam. The decay paths from the $^{16}$C resonances to various states of the final $^{10}$Be or $^{12}$Be nucleus are recognized thanks to the well-resolved $Q$-value spectra. The reconstructed resonances at 16.5(1), 17.3(2), 19.4(1) and 21.6(2) MeV are assigned as the $0^+$, $2^+$, $4^+$ and $6^+$ members, respectively, of the positive-parity $(3/2_\pi^-)^2(1/2_\sigma^-)^2$-bond linear-chain molecular band in $^{16}$C, based on the angular correlation analysis for the 16.5 MeV state and the excellent agreement of decay patterns between the measurements and theoretical predictions. Moreover, another intriguing high-lying state was observed at 27.2(1) MeV which decays almost exclusively to the $\sim$6 MeV states of $^{10}$Be, in line with the newly predicted pure $\sigma$-bond linear-chain configuration. 
\end{abstract}

\maketitle
 
\section{INTRODUCTION}
Nucleon clustering is one of the most intriguing phenomena in nuclear structure studies \cite{Morinaga1956,Ikeda1968,Itagaki2001,Oertzen2006,Horiuchi2012,Freer2018,LiuYang2018,Bijker2020,Ye2020}.
Among the predicted and observed cluster states, chain configuration seems extremely unique, which may even lead to the ring structure \cite{Oertzen2006,Wilkinson1986}.
Studies on chain structure naturally started with the carbon isotopes which may contain three $\alpha$-cluster cores \cite{Morinaga1956,Oertzen2006}. 
Since the chain configuration seems unlikely in $^{12}$C \cite{Itagaki2001}, over the past two decades many theoretical and experimental efforts have been devoted to investigating the molecular-like linear-chain structures in $^{14}$C \cite{Soic2003,Milin2004,Price2007,Haigh2008,Tian2016,Li2017,Zang2018,Yu2021,Freer2014,Fritsch2016,Yamaguchi2017,Suhara2010,Suhara2011,Baba2016,Baba2017,Kanada2020,Yuta2016,Ebran2017} and $^{16}$C \cite{Leask2001,Greenhalgh2002,Bohlen2003,Ashwood2004,DellAquila2016,Maruhn2010,Baba2014,Baba2018,Baba2020}.

The widely adopted antisymmetrized molecular dynamics (AMD) approach has predicted two types of positive-parity liner-chain molecular rotational band in neutron-rich carbon isotopes, associated with the $\pi$-bond and $\sigma$-bond valence neutron configurations, respectively \cite{Suhara2010,Suhara2011,Baba2016,Baba2017,Kanada2020}. 
Most importantly, the AMD calculations have proposed some selective decay patterns from the band members in carbon isotopes to various states of their final decay fragments, based on the structural link property of the decay processes \cite{Baba2018,Baba2017}. This selectivity provides a useful tool to recognize the dominating structure of the mother nucleus via a window associated with the known particular structure of the daughter fragments. For instance, in $^{16}$C, the AMD calculations predicted two positive-parity linear-chain bands with $(3/2^-_\pi)^2(1/2^-_\sigma)^2$ and $(1/2^-_\sigma)^2(1/2_\sigma^+)^2$ valence neutron configurations, respectively \cite{Baba2014,Baba2018,Baba2020}. The former predicted one has the $J^\pi=0^+$ band head at about 16.8 MeV and the moment of inertia ${\hbar^2}/{2\mathfrak{I}}=112$ keV. It is expected that its member states decay into the ground state and the first excited state ($2_1^+$) of $^{12}$Be with specific branching ratios according to the respective Coulomb barriers and the structural links. The same is true for decaying into various states in $^{10}$Be \cite{Baba2014,Baba2018}. On the other hand, the latter predicted band, with a pure $\sigma$-bond configuration, locates at about 15 MeV above the former predicted one and its members decay almost exclusively into the $0^+_2$ state of $^{10}$Be or the 13.6 MeV ($0^+$) state of $^{12}$Be, which possesses also almost pure $\sigma$-bond configuration \cite{Baba2020}. These typical decay strengths can be used to identify the linear-chain configurations in $^{16}$C.

However, clear observation of these selective decay patterns requires precise determination of the low lying states in final fragments. This relies quite often on the resolution of the reaction energy ($Q$-value) spectrum \cite{Li2017,Yu2021}. Unfortunately, the previous experiments aiming at $^{16}$C clustering were not able to achieve this requirement due basically to the broad energy spread of the unstable nucleus beams, detection system performances, and effective statistics  \cite{Leask2001,Greenhalgh2002,Bohlen2003,Ashwood2004,DellAquila2016}.

Following our previous Letter \cite{Liu2020}, we give here an elaborated report of the recent inelastic scattering and cluster decay experiment for $^{16}$C. This measurement aims at the systematic investigation of the selective decay patterns of the high lying resonant states in $^{16}$C, which, in turn, are compared to the latest AMD calculations \cite{Baba2018,Baba2020}. Special efforts were made in detection and data analysis in order to achieve the required high resolutions on the $Q$-value spectra and the relatively high statistics. 

This paper is organized as follows. In Sec. II, the experimental setup and detection technique are described. Section III is dedicated to the data analysis, experimental results and discussions. A brief summary is presented in Sec. IV.

\section{EXPERIMENTAL DETAILS}

\begin{figure*}
    \centering
    \subfigure{\label{fig:Setup}}
    \vspace{0.0in}
    \hspace{0.0in}
    \includegraphics[width=.46\textwidth]{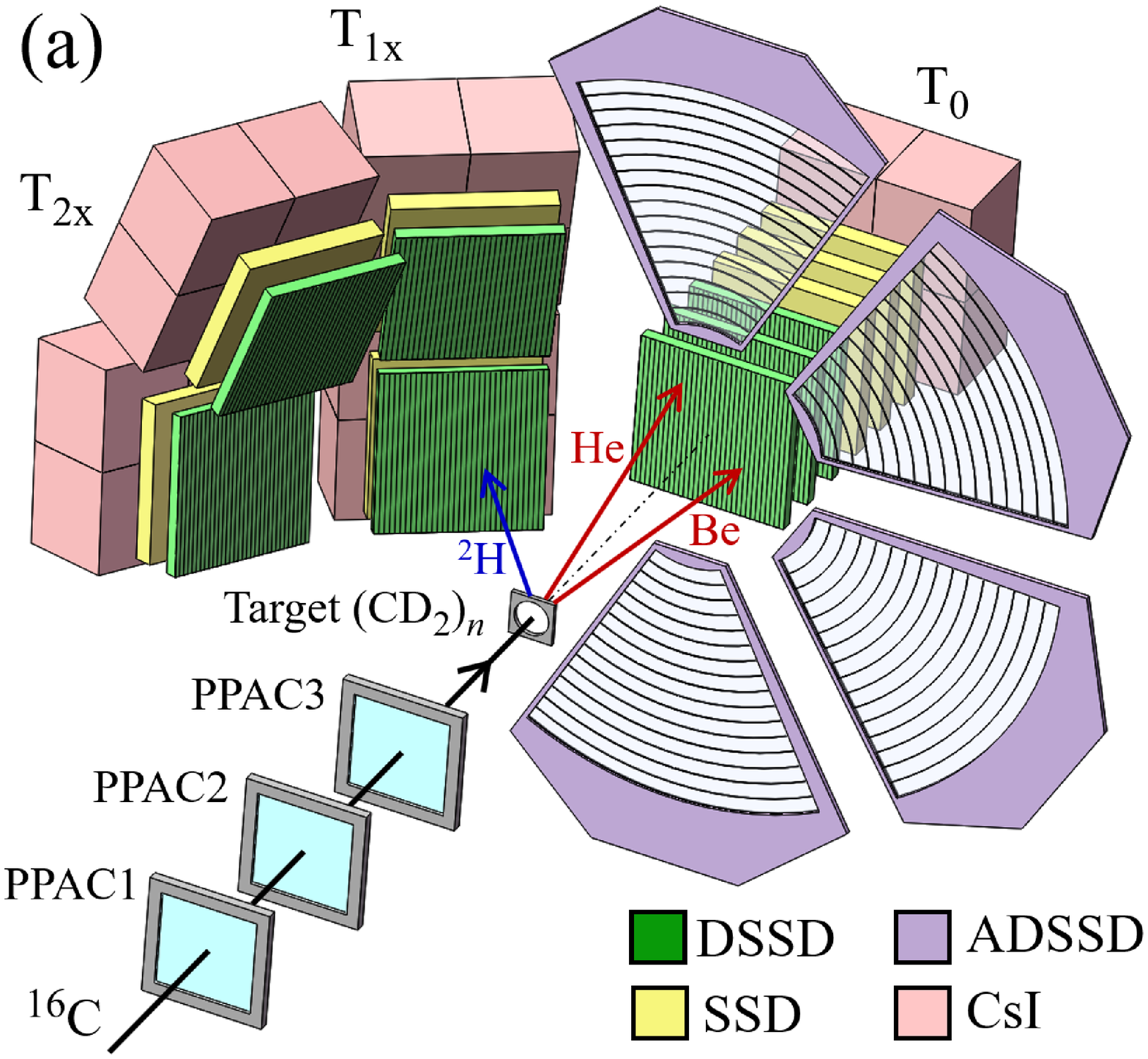}
    \subfigure{\label{fig:BeHeKin}}
    \includegraphics[width=.36\textwidth]{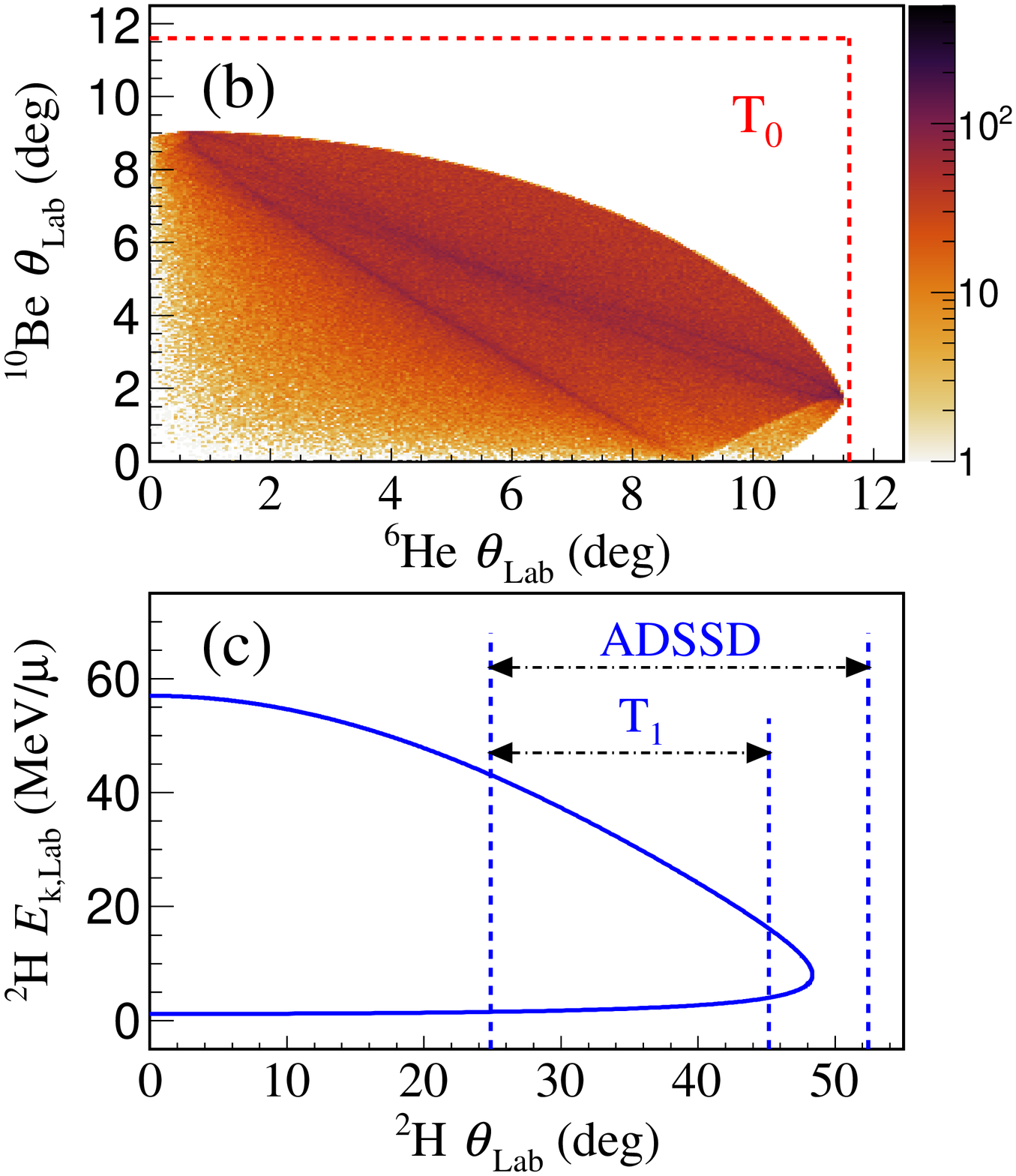}
    \subfigure{\label{fig:HKin}}
    \caption{(a) A schematic layout of the detection system \cite{Liu2020}. T$_\mathrm{1x}$ (T$_\mathrm{2x}$) stands for two telescopes T$_\mathrm{1up}$ and T$_\mathrm{1down}$ (T$_\mathrm{2up}$ and T$_\mathrm{2down}$) arranged at the same polar angle. The products of interest are schematically drawn with arrows in different colors. (b) and (c) Kinematical conditions of the $^2$H($^{16}$C,$^{16}$C$^*$$\rightarrow$$^6$He+$^{10}$Be)$^2$H reaction with $^{16}$C$^*$ in an excited state at 2 MeV above its $^{6}$He decay threshold. The areas within the red-dashed box of (b) and between the blue-dashed lines of (c) depict the angular coverage of the T$_{0}$ and T$_\mathrm{1x}$ telescopes (or ADSSD sectors), respectively.}
    \label{fig:SetupKin}
\end{figure*}

The experiment was carried out at the Radioactive Ion Beam Line at the Heavy Ion Research Facility in Lanzhou (HIRFL-RIBLL) \cite{Sun2003}. A schematic view of the experimental setup is given in Fig. \ref{fig:Setup}. The secondary beam was produced from the fragmentation of a 59.6 MeV/nucleon $^{18}$O primary beam on a 4.5-mm-thick $^9$Be target and identified using the time-of-flight ($\rm TOF$) and energy loss ($\Delta{E}$) measured by two plastic scintillation detectors and a large area single-sided silicon detector installed along the beam line \cite{Sun2003}. The secondary $^{16}$C beam with an energy of $\sim$23.5 MeV/nucleon, a purity of $\sim$90\%, and an intensity of $\sim$1.5$\times$10$^4$ pps was impinged on a 9.53-mg/cm$^2$-thick $(\mathrm{CD}_2)_n$ target. The main contamination of the beam ($\sim10\%$) are the Be, B and C isotopes with the mass/charge number-ratio close to that of $^{16}$C. During the off-line data analysis, pure $^{16}$C isotopes were selected by applying a gate on the TOF-$\Delta E$ spectrum. The beam spot size was about 30 mm in diameter. The beam particles were tracked by three parallel plate avalanche chambers (PPACs) installed upstream of the target, with position resolutions of about 1 mm (FWHM) in both $X$ and $Y$ directions. 

The main reaction channels of interest of this experiment are $^2$H($^{16}$C,$^{16}$C$^*$$\rightarrow$$^4$He+$^{12}$Be)$^2$H and  $^2$H($^{16}$C,$^{16}$C$^*$$\rightarrow$$^6$He+$^{10}$Be)$^2$H in which the target nucleus is much lighter than the projectile. In this case, the emitted nucleus $^{16}$C$^*$ moves in very forward direction, as well as its two decay fragments.
Fig. \ref{fig:BeHeKin} shows the correlated angular distribution for the two decay fragments $^{10}$Be and $^{6}$He in the laboratory system, according to the kinematical calculations in which the mother nucleus $^{16}$C$^*$ is excited to a state at 2 MeV above the corresponding decay threshold.
The shaded strip structures result from the events with $^{16}$C$^{*}$ emitting perpendicularly to the beam axis in the center-of-mass system (c.m.) of the exit channel, or with the decay fragments emitting perpendicularly to the direction of the $^{16}$C$^{*}$ parent. In both cases, the transverse emission corresponds to a certain angular correlation pattern related to the largest probability in the laboratory system. The kinematics for the $^{12}$Be+$^{4}$He decay channel is similar, except for the range of the angular distribution which is slightly larger for $^4$He and smaller for $^{12}$Be. The zero-degree telescope T$_0$, which was located 156 mm downstream from the target covering an angular range of $\sim$(0$\degree$-11.6$\degree$) in the laboratory system (red-dashed box in Fig. \ref{fig:BeHeKin}), could accept almost 100\% of the two decay fragments coincidentally. The T$_0$ array consisted of three 1000-$\mathrm{\mu}$m double-sided silicon strip detectors (DSSD, 64$\times$64 mm$^2$, 32 strips on each side), three 1000-$\mathrm{\mu}$m single-sided silicon detectors (SSD, 64$\times$64 mm$^2$) and a 2$\times$2 CsI(Tl) scintillator array (4.1$\times$4.1$\times$4.0 cm$^3$ for each unit). The three thick DSSDs prevented the majority of the heavier beryllium fragments $^{10}$Be and $^{12}$Be from entering into SSD, while the lighter helium fragments $^6$He and $^4$He were stopped in the subsequent SSD or CsI(Tl) detectors. The fine pixels of DSSDs provided good two-dimensional position resolutions and the capability to record multihit events in one telescope \cite{Qiao2012,Qiao2014}. The inverse kinematics combined with the around zero-degree detection is of high efficiency to measure the particle decay from the near-threshold resonances \cite{Yang20141,Yang20142,Yang2015,Zang2018,Liu2020,LiuWei2020,LiuWei2021}. Moreover, the beam particles which did not react with the target were stopped in the first two layers of the T$_0$ telescope with their energies being measured precisely. Thus, an energy spread of 10.2 MeV (FWHM) was obtained for the $^{16}$C secondary beam. Such a large uncertainty for the projectile-fragmentation (PF) secondary beam is the main reason for the poor resolutions of the normal $Q$-value spectra deduced from two-fold coincident events, which will be addressed later. 

The recoil $^2$H was measured by four other Si-CsI telescopes (T$_\mathrm{1x}$ and T$_\mathrm{2x}$) and four sectors of the annular double-sided silicon strip detectors (ADSSD). The T$_\mathrm{1x}$ and T$_\mathrm{2x}$ telescopes were placed at 178.7 and 156.6 mm from the target, and covered an angular range of $\sim$(24.8$\degree$-45.2$\degree$) and $\sim$(24.9$\degree$-52.4$\degree$) in the laboratory system, respectively. Each of them comprised a thin DSSD (60 $\mathrm{\mu}$m for T$_\mathrm{1x}$, and 300 $\mathrm{\mu}$m for T$_\mathrm{2x}$), a thick DSSD (1500 $\mathrm{\mu}$m), and a 2×2 CsI(Tl) scintillator array. T$_\mathrm{1x}$ accepted part of the recoil $^2$H, as shown in Fig. \ref{fig:HKin}. When the $^{16}$C$^*$ was excited to a lower state, the distribution of the recoil $^2$H would extend to the angular range covered by T$_\mathrm{2x}$. 
One 150-$\mathrm{\mu}$m-thick and three 400-$\mathrm{\mu}$m-thick sectors of ADSSD were located around the $\rm{T_0}$ telescope and at a distance of 123 mm from the target. Each sector has an inner (outer) radius of 32.6 mm (135.1 mm) covering an angular range of $\sim$(24.9$\degree$-52.4$\degree$) in the laboratory system\cite{Chen2018}. The front side is divided into sixteen 6.4-mm-wide ring strips, while the back side eight wedge-shaped regions. So, these sectors accepted part of the recoil $^2$H particles, as depicted in Fig. \ref{fig:HKin}, and provided the deposited energies and the track positions of these particles.
All of these $\rm{T_{0}}$, $\rm{T_{1x}}$ and $\rm{T_{2x}}$ telescopes and ADSSD sectors were installed in a compact structure in order to cover a larger solid angle for the three-fold coincident detection.

\begin{figure}
    \centering
    \includegraphics[width=.45\textwidth]{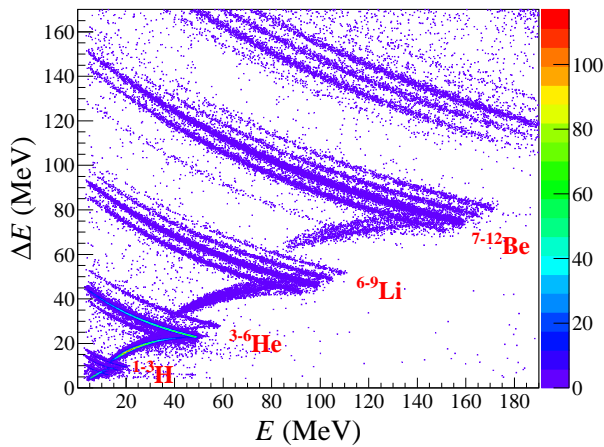}
    \caption{Particle identification (PID) spectrum measured by the first two layers of the $\rm{T_0}$ telescope using the $\Delta E$-$E$ method. Note that $^5$He and $^8$Be are missing from the observed bands as expected.}
    \label{fig:PID}
\end{figure}

The overall energy match for different strips in one DSSD was achieved according to the self-uniform calibration method as described in Ref. \cite{Qiao2014}.
And then the absolute energy calibration of each silicon-detector was accomplished by using a combination of $\alpha$-particle sources and following the procedures described in Refs. \cite{Qiao2014,Liu2018,Tao2019,Manfredi2018}.
The characteristic energy resolutions of the present silicon detectors are about 1\% for the 5.486-MeV $\alpha$ particles emitted from the $\rm {^{241}Am}$ source.
The energy calibration for CsI(Tl) scintillators was realized by the procedures described in Refs. \cite{Wagner2001,DellAquila2019,Li2021}.
Although some nonlinear responses for the CsI(Tl) light output have been reported in the literature, a linear formula can be a good approximation for light ions like He isotopes for the present measurements \cite{Freer2001}.
Timing information obtained from the beam monitors and the DSSD strips was applied to exclude most accidentally coincident signals. This is of particular importance for the $\rm{T_0}$ telescope which was directly exposed to the beam at a hitting-rate higher than $10^4$ Hz \cite{Zang2018,Liu2020}. Particles produced from the nuclear reactions on the detector layers, but not on the physics target, were largely eliminated by the tracking analysis combining hit positions on the target and several neighboring DSSD layers \cite{Zang2018,Yang20191,Yang20192}.
Thanks to the excellent energy, timing, and position resolutions of the telescopes, isotopes from hydrogen to beryllium were unambiguously identified based on the standard energy loss versus residual energy ($\Delta E-E$) technique, as displayed in Fig. \ref{fig:PID}. This particle identification (PID) was also confirmed by the simulation results using energy-loss tables \cite{Ziegler2010}.
The energy losses of all initial and final particles in the target were corrected according to the energy-loss tables \cite{Ziegler2010}, under the assumption that the reaction point is at the center of the target along the beam direction. This correction, about 3.9 MeV for the $^{16}$C beam particle for instance, is essential in order to obtain a correct and well resolved $Q$-value spectrum (see section \ref{chap:kin}).

This experiment focused on measuring the decay of $^{16}$C at forward angles, with the detection of at least two decay fragments in the T$_0$ telescope. Therefore the main trigger of the detection system was set for at least two hits in coincidence on each of the front side of the first two DSSDs in T$_0$. Unfortunately, these events include those inter-strip hitting by one particle which generates two signals at adjacent strips. In fact these one-particle inter-strip hitting events may be much more than the events with two particles hitting two adjacent strips, especially for the in-beam detection such as that with our T$_0$ telescope. In previous experiments \cite{Yang20141,Yang20142,Yang2015,Zang2018}, events with signals on two adjacent strips were discarded for simplicity at the expense of detection efficiency particularly for resonances close to the decay threshold. In this work, we have made extensive efforts to analyze these adjacent-hitting events in order to recover as much as possible the near-threshold detection efficiency. The basic technique is to check the adjacent signals by matching the energies from  both sides of one DSSD as well as the hitting positions of neighboring DSSD layers, together with the location of the related energy pairs on the respective PID ($\Delta E-E$) spectra. 
As a result, the statistics of the excitation energy spectra of $^{16}$C, especially at the regions near the cluster decay thresholds, have been significantly improved compared to the previous analysis \cite{Yang20141,Yang20142,Yang2015,Zang2018}. 

\begin{figure}
    \centering
    \subfigure{\label{fig:Be8C12IMa}}
    \vspace{-0.16in}
    \hspace{-0.0in}
    \subfigure{\label{fig:Be8C12IMb}}
    \includegraphics[width=.45\textwidth]{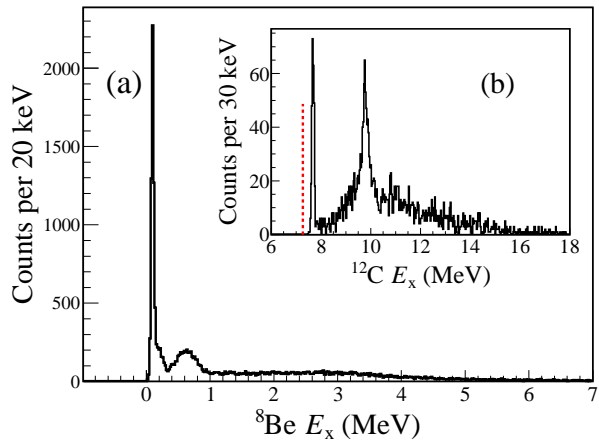}
    \caption{Excitation energy spectra reconstructed from  (a) 2-$\alpha$ and (b) 3-$\alpha$ coincident events. The red dotted line in (b) stands for the 3-$\alpha$ decay threshold (7.27 MeV) of $^{12}$C.}
    \label{fig:Be8C12IM}
\end{figure}

The detection and calibration were validated by using the two- and three-$\alpha$ coincident events to reconstruct the known $\rm{^8Be}$ and $\rm{^{12}C}$ resonances, respectively \cite{Wuosmaa1992,DellAquila2016,Smith2017}, as exhibited in Fig. \ref{fig:Be8C12IM}.
Here $^8$Be and $^{12}$C are the intermediate fragments produced by $^{16}$C breakup \cite{Ashwood2004,DellAquila2016}.
Using the 2-$\alpha$ events,  we see in Fig. \ref{fig:Be8C12IMa} a narrow peak at about 91.8 keV and a broad one at around 3 MeV, which are consistent with the $\alpha$-$\alpha$ decay of $^8$Be in its ground and the first excited state, respectively. In addition, a ghost peak appears in the vicinity of 600 keV owing to the neutron decay of the second excited state of $^9$Be (2.43 MeV, $5/2^-$) \cite{Ashwood2004,DellAquila2016}.
Using the 3-$\alpha$ events, it appears in Fig. \ref{fig:Be8C12IMb} the prominent Hoyle state (7.654 MeV, $0^+$) of $^{12}$C \cite{Smith2017}, with a narrow width (FWHM) of only 0.108 MeV. Furthermore, the 9.64 MeV ($3_1^-$) excited state in $^{12}$C is also strongly populated with a width (FWHM) of 0.172 MeV, indicating an energy resolution around this peak of about 0.167 MeV (FWHM) since the intrinsic width of this state is very small ($\sim$ 42 keV) \cite{Freer2009}.

\section{RESULTS AND DISCUSSION}
\subsection{Kinematics Reconstruction Methods} \label{chap:kin}
For a reaction and decay process \text{a(A, B$^*$$\rightarrow$C+c)b}, where an inelastic scattering or transfer reaction between the projectile A and the target nucleus a is followed by a prompt decay of the resonant nucleus B$^*$ into two fragments C and c accompanied by the recoil particle b, the conservation of energy and momentum can be expressed as 
\begin{align}
    E_{\mathrm{A,Lab}} + E_{\mathrm{a,Lab}} &= E_{\mathrm{C,Lab}} + E_{\mathrm{c,Lab}} + E_{\mathrm{b,Lab}}, \label{eq:EConservatin1} \\
    \boldsymbol{p}_{\mathrm{A,Lab}} + \boldsymbol{p}_{\mathrm{a,Lab}} &= \boldsymbol{p}_{\mathrm{C,Lab}} + \boldsymbol{p}_{\mathrm{c,Lab}} + \boldsymbol{p}_{\mathrm{b,Lab}}, \label{eq:PConservation1}
\end{align}
where $E_{i,\mathrm{Lab}}$ and $\boldsymbol{p}_{i,\mathrm{Lab}}$ $(i=\mathrm{A,a,C,c,b})$ are the relativistic total energy and the momentum of each particle in the laboratory system, respectively. The relationships among these two quantities and the kinetic energy $T_{i,\mathrm{Lab}}$ for each particle are given by
\begin{align}
    E_{i,\mathrm{Lab}}^2 - \boldsymbol{p}_{i,\mathrm{Lab}}^2c^2 = m_{i0}^2c^4, \label{eq:EPRelation} \\
    E_{i,\mathrm{Lab}} = T_{i,\mathrm{Lab}} + m_{i0}c^2, \label{eq:ETRelation}
\end{align}
where $m_{i0}$ $(i=\mathrm{A,a,B,C,c,b})$ is the rest mass for each particle. 
For the present experiment, the projectile A ($^{16}$C), target a ($^{2}$H), and the detected recoil particle b ($^{2}$H) and decay fragment c ($^{4}$He or $^{6}$He) are all in their ground states, while the other decay fragment C ($^{12}$Be or $^{10}$Be) may be in its ground or bound excited state.
The rest mass of each particle in its ground state can be found from the standard nuclear date sheet with high precision.
Assuming that the target is at rest and the kinetic energies and directions of any three of the other four particles (i.e. A, C, c and b) were measured, those for the undetected particle together with the mass of the decay fragment C in a certain state can be deduced. 
We note that the bound excited C will normally decay to the ground state by emitting a photon before being detected. In the laboratory system, the momentum taken away by the photon is negligible in comparison to that of the massive nucleus \cite{David2020}. 
As a result, the reaction $Q$-value, which is defined by the mass deficit between the initial and final particles, can be calculated from the energy released in the reaction process:
\begin{align}
    Q &= (m_\mathrm{A0} + m_\mathrm{a0} - m_\mathrm{C0} - m_\mathrm{c0} - m_\mathrm{b0})c^2 \notag \\
      &= T_{\mathrm{C,Lab}} + T_{\mathrm{c,Lab}} + T_{\mathrm{b,Lab}} - T_{\mathrm{A,Lab}} \notag \\
      &= Q_{\mathrm{ggg}} - E_{\mathrm{x,C}}, \label{eq:Q}
\end{align}
where $E_{\mathrm{x,C}}$ is the excitation energy of the decay fragment C, and $Q_{\mathrm{ggg}}$ stands for the $Q$-value of the reaction with all final particles in their ground states (g.s.).

In the present experiment, the incident directions of the $^{16}$C projectiles were determined by the PPAC detectors event-by-event, and the mean kinetic energy of the $^{16}$C beam can be given by the T$_0$ telescope. The trigger setting of the data acquisition (DAQ) system allowed coincident detection of two decay fragments by the forward T$_0$ telescope. In the meantime, the recoil $^2$H particles were also passively recorded by the T$_\mathrm{1x}$, T$_\mathrm{2x}$ telescopes and ADSSD sectors. 
Previously, in a standard two-fold coincident method, only the two decay fragments C and c were measured and the energy of the recoil particle b could be obtained by using momentum conservation between A and C+c+b \cite{Zang2018}. In this case, the $Q$-value resolution was often limited by the large energy spread of the secondary-beam particle A.  On the other hand, when all three final particles C+c+b are measured (three-fold coincident method), the energy of the projectile A can be deduced event-by-event through the momentum conservation. In this case, the $Q$-value resolution is purely relying on the detection system. These two methods will be compared in the following.

Furthermore, the mass of the intermediate mother nucleus B$^*$ can be reconstructed from the decay process B$^*$$\rightarrow$C+c. The conservation of energy and momentum gives
\begin{align}
    E_{\mathrm{B,Lab}} &= E_{\mathrm{C,Lab}} + E_{\mathrm{c,Lab}}, \label{eq:EConservatin2} \\
    \boldsymbol{p}_{\mathrm{B,Lab}} &= \boldsymbol{p}_{\mathrm{C,Lab}} + \boldsymbol{p}_{\mathrm{c,Lab}}, \label{eq:PConservation2}
\end{align}
where $E_{i,\mathrm{Lab}}$ and $\boldsymbol{p}_{i,\mathrm{Lab}}$ $(i=\mathrm{B,C,c})$ are the relativistic total energy and the momentum of each particle in the laboratory system, respectively. The rest mass of B$^*$ can then be deduced according to equation \eqref{eq:EConservatin2} and \eqref{eq:PConservation2}, together with definitions \eqref{eq:EPRelation} and \eqref{eq:ETRelation}. 
Therefore, the decay $Q$-value or the relative energy ($E_\mathrm{rel}$) of the two fragments for the decay process can be expressed as:
\begin{align}
    Q_\mathrm{decay} = E_\mathrm{rel} = (m_\mathrm{B^{*}0} - m_\mathrm{C0} - m_\mathrm{c0})c^2. \label{eq:Q2}
\end{align}
The excitation energy of the resonant state B$^{*}$ is then
\begin{align}
    E_\mathrm{x} &= (m_\mathrm{B^{*}0} - m_\mathrm{B_{gs}0})c^2 \notag \\
                 &= E_\mathrm{rel} + E_\mathrm{th}, \label{eq:Ex}
\end{align}
with $E_{\mathrm{th}}=m_\mathrm{C0}+m_\mathrm{c0}-m_\mathrm{B_{gs}0}$ the threshold energy (or separation energy) of the current decay channel.
This method is known as the invariant mass method and has been widely used in our previous works  \cite{Jiang2017,Feng2019,Jiang2020,Li2017,Yang20191,Yang20192,Yu2021}.

\subsection{$Q$-value spectra}

\begin{figure}
    \centering
    \subfigure{\label{fig:Qa}}
    \vspace{-0.16in}
    \hspace{-0.0in}
    \subfigure{\label{fig:Qb}}
    \vspace{-0.0in}
    \hspace{-0.0in}
    \subfigure{\label{fig:Qc}}
    \vspace{0.0in}
    \hspace{0.0in}
    \subfigure{\label{fig:Qd}}
    \includegraphics[width=.45\textwidth]{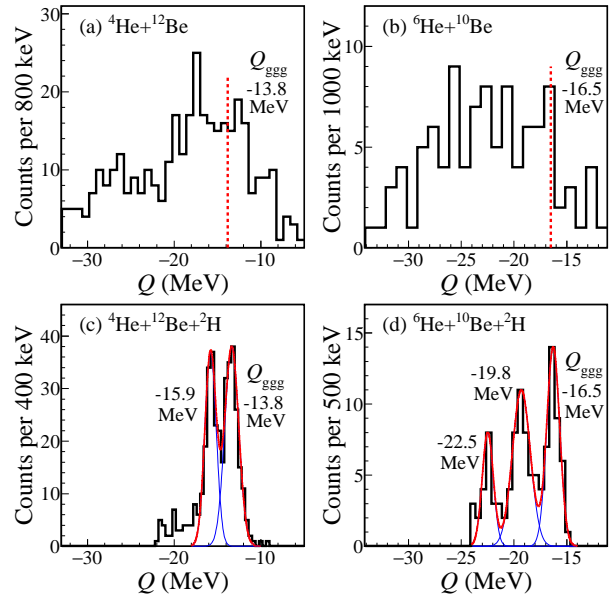}
    \caption{$Q$-value spectra for  (a,c) $^{16}$C$\rightarrow^{4}$He+$^{12}$Be and (b,d) $^{16}$C$\rightarrow^{6}$He+$^{10}$Be breakup reactions on a $^2$H target. The upper and lower panels were obtained from the two-fold and three-fold coincident analysis, respectively. In the lower panels, $Q$-value peaks for decaying into $^{12}$Be($0^+_1$), $^{12}$Be($2^+_1$), $^{10}$Be($0^+_1$), $^{10}$Be($2^+_1$) and $^{10}$Be($\sim$6 MeV) are fitted by the Gaussian functions (blue-thin lines for the peaks and red-thick line for their sum).}
    \label{fig:Q}
\end{figure}

As mentioned above, the resolution of reaction $Q$-value is of essential importance to discriminate various decay paths. The $Q$-value spectra obtained in this experiment are shown in Fig. \ref{fig:Q}. For the sake of comparison, the same three-fold coincident event sample was used although different $Q$-value deduction methods can be applied to these events. Firstly, we may apply the general two-fold coincident method, in which the momentum of the recoil particle, assuming $^2$H, was deduced according to the momentum conservation by using the momentum of the beam and the detected two forward fragments, as discussed in section \ref{chap:kin}. Because of the relatively large energy spread of the PF-type radioactive ion beam and the inaccurate energy monitoring over the very long secondary beam line, the extracted $Q$-value spectra can hardly reach the required resolution, as can be seen in Fig. \ref{fig:Q}(a)(b) and also demonstrated previously \cite{Zang2018}. In addition, if we take the real two-fold event sample, without requiring the recoil-deuteron identification in surrounding detectors, the breakup of the weakly-bound $^2$H would further contaminate the  $Q$-value spectra, particularly by creating a long tail in the lower $Q$-value side (not shown in the figure). This has motivated us to pursue the real three-fold coincident detection and analysis in order to obtain clear decay patterns for the $^{16}$C resonances.  

Taking advantages of the triple coincident detection in which all final particles were identified without ambiguity, the beam energy can be deduced event-by-event according to the momentum conservation, as discussed in section \ref{chap:kin}. Hence, the $Q$-value resolution relies solely on the performance of the detection system, but not on the beam energy monitoring. For the first time, in PF-type experiments, $Q$-value peaks corresponding to the ground and low-lying excited states of the final fragments are clearly discriminated, as demonstrated in Fig. \ref{fig:Q}(c)(d). The spectra are fitted with Gaussian peaks. For $^4$He + $^{12}$Be decay channel [Fig. \ref{fig:Qc}], the $Q_{\rm{ggg}}$ peak at about -13.8 MeV corresponds to the reaction for all final particles in their ground states. 
Another peak at about -15.9 MeV is mainly associated with $^{12}$Be in its $2^+_1$ (2.109 MeV) state.
The decay to the nearby $0_2^+$ (2.251 MeV) state should be much weaker because of the much smaller penetration factors in comparison to those for the $2^+_1$ (2.109 MeV) state.
The decay to another nearby $1_1^-$ (2.715 MeV) state of $^{12}$Be is also unlikely because it should stand at the far edge of the actual $Q$-value peak but apparently no structure appears there.
For $^6$He + $^{10}$Be decay channel [Fig. \ref{fig:Qd}], the highest peak at about -16.5 MeV is for the $Q_\mathrm{ggg}$, and another two at about -19.8 MeV and -22.5 MeV are associated with $^{10}$Be in its first excited state ($2_1^+$, 3.368 MeV) and the four adjacent states at around $\sim$6 MeV ($2_2^+$, $1_1^-$, $0_2^+$, $2_1^-$), respectively. Owing to the much higher energy of the first excited state in $^4$He (20.21 MeV), and inexistence of bound excited states in $^6$He and $^2$H, all observed peaks cannot correspond to the excitation of $\rm{^4He}$, $\rm{^6He}$ or $^2$H. In the following analysis of the $^{16}$C resonances, we use only three-fold coincident events and always apply gates on certain $Q$-value peaks to select the decay paths.

The contamination in the $Q$-value spectra from the proton and carbon components in the ($\rm{CD_2}$)$_n$ target, as well as the target holder, has been estimated to be negligible for the triple coincident events, according to the analysis for empty and pure-C target runs. This demonstrates that the three-fold coincident measurement is sufficiently effective to reduce the background contamination, even for detection covering the beam direction.

\begin{figure}
    \centering
    \subfigure{\label{fig:IMa}}
    \vspace{-0.16in}
    \hspace{-0.0in}
    \subfigure{\label{fig:IMb}}
    \vspace{-0.0in}
    \hspace{-0.0in}
    \subfigure{\label{fig:IMc}}
    \vspace{0.0in}
    \hspace{0.0in}
    \subfigure{\label{fig:IMd}}
    \vspace{-0.0in}
    \hspace{-0.0in}
    \subfigure{\label{fig:IMe}}
    \includegraphics[width=.45\textwidth]{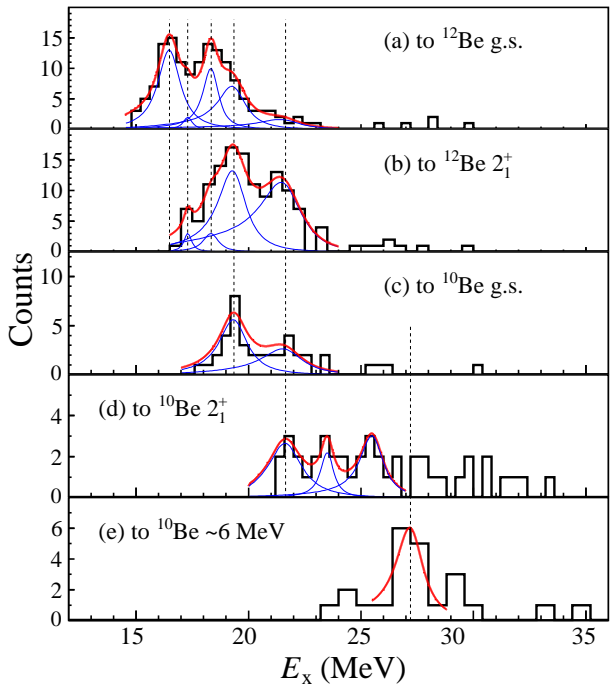}
    \caption{Excitation energy spectra of $^{16}$C reconstructed from two final channels ($^4$He + $^{12}$Be + $^2$H and $^6$He+$^{10}$Be + $^2$H) after gating on various $Q$-peaks in Fig. \ref{fig:Qc} and \ref{fig:Qd}. Each spectrum is fitted with  the sum (red-thick line) of several peaks (blue-thin lines). The vertical black dashed lines are plotted to guide the eyes for the corresponding states. The details are described in the text.}
    \label{fig:IM}
\end{figure}

\subsection{Excitation-energy spectra}

\begin{figure}
    \centering
    \includegraphics[width=.43\textwidth]{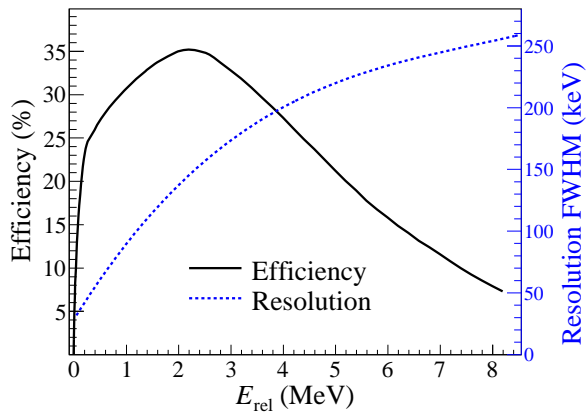}
    \caption{Simulated detection efficiency (black-solid line) and resolution (blue-dotted line) as a function of the relative energy for $^{16}$C decaying into $^4$He + $^{12}$Be(g.s.).}
    \label{fig:ResEff}
\end{figure}

After the selection of the decay paths by gating on the $Q$-value peaks [Fig. \ref{fig:Q}(c)(d)] corresponding to various states of the final  nucleus $^{12}$Be or $^{10}$Be, the excitation energies of $^{16}$C for the decay process of $^4$He + $^{12}$Be [Fig. \ref{fig:IM}(a)(b)] and $^6$He + $^{10}$Be [Fig. \ref{fig:IM}(c)-(e)] were reconstructed according to the invariant mass method as described in section \ref{chap:kin}. Mont Carlo simulations were performed to evaluate the detection efficiency of the triple coincident events (black-solid line in Fig. \ref{fig:ResEff}). The simulation took into account the energy spread and angular straggling of the beam particles, the reaction position and energy loss in the target, the real experimental setup, and the energy and position resolutions of the detectors. The angular distribution of the inelastic scattering process was presumed to follow an exponential form, while that of the cluster decay process was taken as isotropic in the center-of-mass system \cite{Li2017,DellAquila2016}. The zero-degree telescope (T$_0$) employed in the present measurement, with fine pixels and an angular coverage focusing on the most forward angles, offered a remarkably high efficiency for the measurement of near-threshold resonant states. The detection efficiency for $^{16}$C decaying into the $^4$He + $^{12}$Be(g.s.) channel is shown in Fig. \ref{fig:ResEff}, with a maximum value of $\sim$ 35\% at small relative energy ($E_\mathrm{rel}$), which decreases to $\sim$ 8\% at $E_\mathrm{rel} = 8$ MeV. Similar detection efficiencies were also found for the other decay channels, such as $^4$He + $^{12}$Be$(2_1^+)$, $^6$He + $^{10}$Be(g.s.), $^6$He + $^{10}$Be$(2_1^+)$, and $^6$He + $^{10}$Be($\sim$6 MeV). The detection efficiency is mainly limited by the angular coverage of the $\rm{T_{1x}}$, $\rm{T_{2x}}$ telescopes and ADSSD sectors. However, the drop down close to the threshold and the decline at high excitation energy are mainly due to the configuration and coverage of the DSSDs in the $\rm{T_0}$ telescope. When the relative energy of the two fragments ($E_\mathrm{rel}$) is close to zero, they will mostly hit the same strip of DSSD and deteriorate the particles identification becomes. On the other hand, when $E_\mathrm{rel}$ is too large, one or both of the fragments may escape from the array with appreciable probabilities. The resolution of the reconstructed excitation (or relative) energy can also be  estimated according to the Mont  Carlo  simulations \cite{Yang20141,Yang20142,Yang2015,Zang2018}, as plotted in Fig. \ref{fig:ResEff} (blue-dotted line). For $^{16}$C decaying into the $^4$He + $^{12}$Be(g.s.) channel, the resolution (FWHM) is determined to be around 90 keV at $E_\mathrm{rel}=1$ MeV and increases to about 250 keV at $E_\mathrm{rel} = 8$ MeV.

The excitation energy spectra in Fig. \ref{fig:IM} were fitted with several resonant peaks (Breit-Wigner form),  each convoluted with energy-resolution functions (Gaussian form) and modified by the detection efficiencies \cite{Schiller2007,Cao2012,Tanaka2017}. The initialization of the parameters was chosen according to the visual observation of possible peaks in various spectrum panels (Fig. \ref{fig:IM}) and referring to the previous tentative assignments \cite{DellAquila2016}. In the fitting procedure, the centers and widths of each resonance were kept consistent for various decay paths. The extracted positions and the physical widths of the populated resonant states are listed in Table \ref{table:Levels}. The AMD predictions from Ref. \cite{Baba2018} are also presented for comparison which are updated by the AMD calculations from Ref. \cite{Baba2020}. The five peaks at 16.5(1), 17.3(2), 18.3(1), 19.4(1), and 21.6(2) MeV are observed for the first time in this experiment, which will be discussed in details below. The errors in the parenthesis are statistical only. In addition, a systematic uncertainty of about 100 keV can be estimated for each peak position due basically to the detection performances \cite{Li2017}. It is worth noting in Table \ref{table:Levels} that the previously reported peak at about 20.6 MeV in Ref. \cite{DellAquila2016} is not observed in this measurement. In that work, this peak was reconstructed from the $^6$He + $^{10}$Be channel without $Q$-value selection and the rest mass of $^{10}$Be in the ground state was simply used regardless of its possible excitation. According to the invariant mass method described in Eqs. \eqref{eq:EConservatin2} and \eqref{eq:PConservation2}, the latter would cause a shift of the reconstructed energy. For instance, the presently observed 23.5 MeV peak in Fig. \ref{fig:IMd} or 27.2 MeV peak in Fig. \ref{fig:IMe} could erroneously be shifted to about 20.6 MeV in Fig. \ref{fig:IMc} if the the $Q$-value selection was not realized.

\begin{table}
\caption{Presently measured excitation energies, spin parities, and total decay widths of the resonances in $^{16}$C, in comparison to those from the previous experiment \cite{DellAquila2016} and AMD calculations \cite{Baba2018,Baba2020}. The Errors in parentheses for positions and widths of the observed resonances are only statistical.}
\label{table:Levels}
\begin{threeparttable}
\begin{ruledtabular}
\begin{tabular}{ c c c c c c }
\multicolumn{3}{c}{This work} & \multicolumn{1}{c}{Exp. \cite{DellAquila2016}} & \multicolumn{2}{c}{AMD \cite{Baba2018,Baba2020}} \\
\cmidrule(r){1-3}  \cmidrule(r){5-6}
$E_\mathrm{x}$(MeV) & $J^{\pi}$ & $\Gamma_\mathrm{tot}$(keV) & $E_\mathrm{x}$(MeV) & $E_\mathrm{x}$(MeV) & $J^{\pi}$ \\
\midrule
16.5(1) & 0$^+$ & 1200(200) &      & 16.81\tnote{\textdagger} & $0_{6}^+$  \\
17.3(2) &       & 400(200)  &      & 17.51\tnote{\textdagger} & $2_{9}^+$  \\
18.3(1) &       & 800(100)  &      &       &            \\
19.4(1) &       & 1500(160) &      & 18.99\tnote{\textdagger} & $4_{10}^+$ \\
        &       &           & 20.6\tnote{*} &       &            \\
21.6(2) &       & 2200(200) &      & 21.49\tnote{\textdagger} & $6_{5}^+$  \\
23.5(2) &       & 680(200)  &      &       &            \\
25.5(2) &       & 1230(200) &      &       &            \\
27.2(1) &       & 1460(200) &      & 31.72\tnote{\textdaggerdbl} & $0^+$      \\
        &       &           &      & 31.98\tnote{\textdaggerdbl} & $2^+$      \\
        &       &           &      & 32.72\tnote{\textdaggerdbl} & $4^+$      \\
\end{tabular}
\begin{tablenotes}
\footnotesize
\item[*] state without $Q$-value selection
\item[\textdagger] states reported in Ref. \cite{Baba2018}
\item[\textdaggerdbl] states reported in Ref. \cite{Baba2020}
\end{tablenotes}
\end{ruledtabular}
\end{threeparttable}
\end{table}

\subsection{Angular correlation}
The latest AMD calculations \cite{Baba2018} have proposed a positive-parity linear-chain molecular band headed by the 16.81 MeV ($0^+$) state which is close to the presently observed 16.5 MeV state [Fig. \ref{fig:IMa} and Table \ref{table:Levels}]. This reconstructed peak is a good candidate for angular correlation analysis, owing to its clear peak identification, the highest statistics and few background contamination.  

The angular correlation between the decay fragments is a sensitive tool to determine the spin of the mother resonant state, which has been successfully applied in the previous works \cite{Freer1996,Freer1999,Curtis2001,Yang20141,Yang20142,Yang2015,Yang20191,Yang20192}. This correlation is based on the angular distribution of the decay fragments in the c.m. of the mother nucleus, and thus independent on the population mechanisms. The angles involved here are $\theta^*$, the c.m. scattering polar angle of the mother nucleus in its resonant state, and $\psi$, the polar angle of the relative velocity vector between the two decay fragments. For a resonant state with an angular momentum $J$, which subsequently decays into two spin-0 fragments, the angular correlation spectrum is proportional to a Legendre polynomial of order $J$, $|P_J(\mathrm{cos}(\psi))|^2$. In the case of $J > 0$ and the detection away from $\theta^*=0\degree$, the correlation pattern will deviate from the ideal $|P_J(\mathrm{cos}(\psi))|^2$ distribution. This effect can be corrected by modifying $\psi$ with a phase shift factor $a=-\frac{l_i-J}{J}$ being used for the projection onto the $\theta^*=0$ axis \cite{Freer1996,Freer1999}. Here $l_i$ is the dominant partial wave in the entrance channel assuming surface collision, namely $l_i=r_0(A_p^{1/3}+A_t^{1/3})\sqrt{2\mu E_{\rm{c.m.}}}$, where $A_p$ and $A_t$ are mass numbers of the beam and the target nucleus, respectively, with $\mu$ the reduced mass and $E_{\rm{c.m.}}$ the c.m. energy. However, this correction is not necessary for $J=0$, as it corresponds to the isotropic decay in the c.m. frame of the resonant state \cite{Yang20141,Yang20142,Yang2015}. Owing to the symmetry property of $|P_J(\mathrm{cos}(\psi)|^2$ function and the uniform behaviour of the experimental distribution over the whole range of cos$\psi$, the analysis is performed against $|\mathrm{cos}(\psi)|$ only, in order to have a better statistical presentation \cite{DellAquila2016,Yang20191,Yang20192}.

\begin{figure}
    \centering
    \includegraphics[width=.45\textwidth]{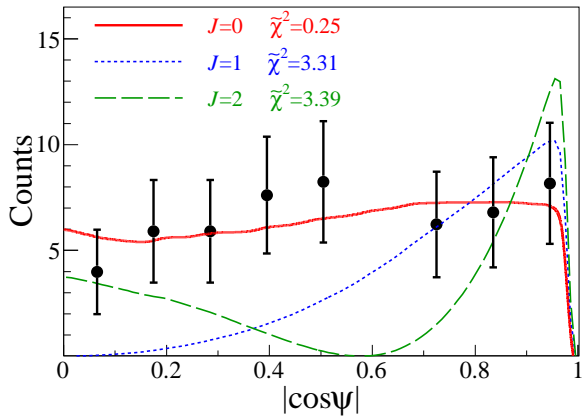}
    \caption{Angular correlation between the $^4$He and $^{12}$Be decay fragments from the 16.5 MeV resonance in $^{16}$C. Experimental results (black-solid circles) are compared with the Legendre polynomials of order 0 (red-solid line), 1 (blue-dotted line) and 2  (green-dashed line) corrected by the detection efficiency. The corresponding reduced $\chi^2$ values are also indicated.}
    \label{fig:Spin0}
\end{figure}

\begin{figure}
    \centering
    \includegraphics[width=.45\textwidth]{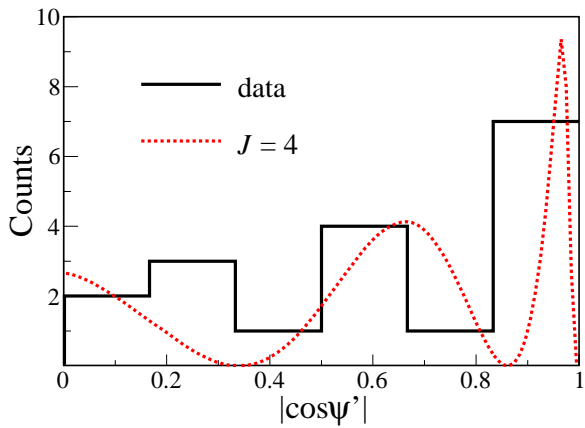}
    \caption{Angular correlation between the $^6$He and $^{10}$Be decay fragments from the 19.4 MeV resonance in $^{16}$C. The present experimental data (histogram) are projected onto $\theta^*$ = 0 axis assuming spin 4, and compared to the Legendre polynomial of order 4 (red-dotted line) corrected by the detection efficiency.}
    \label{fig:Spin4}
\end{figure}

The spectrum such as Fig. \ref{fig:IMa} can be plotted against each selected $|\mathrm{cos}(\psi)|$ bin and the counting number in that bin for certain resonant states can then be extracted from the same peak fitting as for the summed spectrum \cite{DellAquila2016,Yang20191,Yang20192}. In this fitting process, only the amplitude of each peak was left as free parameters, while the position and width were fixed the same as for the summed spectrum.
The currently extracted angular correlation data for the 16.5 MeV state are plotted in Fig. \ref{fig:Spin0} and compared with the theoretical $|P_J(\mathrm{cos}(\psi))|^2$ distributions assuming $J$ values of 0, 1 and 2. In the calculations the detection efficiency has been taken into account. The loss of events in the experimental distribution and the drop down of the theoretical curves, at $|\mathrm{cos}(\psi)|\sim 1$, is attributed to the ineffectiveness of the coincident measurement for the two close-by fragments entering into the same DSSD-strip of the $\mathrm{T_0}$ telescope. As evidenced in Fig. \ref{fig:Spin0}, the spin-0 component is consistent with the data, whereas spin 1 and spin 2 (and higher spins with more oscillations) can be excluded due basically to the behavior at the minima and also to the much larger reduced $\chi^2$ values.
Since the 16.5 MeV state is a relatively well isolated peak [Fig. \ref{fig:IMa}], a simple cut can also be applied instead of a fitting to extract the count number. This was used in our previous work \cite{Liu2020}, where similar shape of the correlation spectrum was reported. The statistically consistent results from various extracting methods further confirm the present spin assignment for the 16.5 MeV state. Consequently, the observed 16.5 MeV state can be considered as the most promising candidate for the $0^+$ band head of the positive-parity linear-chain rotational band of $^{16}$C.

As a cross-check, we tried also the standard angular correlation analysis \cite{DellAquila2016,Yang20191,Yang20192} for the observed 19.4 MeV state which is quite isolated in the decay path to $^{10}$Be(g.s.) [Fig. \ref{fig:IMc} and Table \ref{table:Levels}]. In the present work, we take $r_0=1.4$ fm, which leads to $l_i=10\hbar$. As can be seen in Fig. \ref{fig:Spin4}, despite of the low statistics, the oscillatory behavior of the experimental spectrum does not exclude a spin-4 assignment for the 19.4 MeV state, and therefore not contradict the following decay-pattern analysis. The same kind of analyses were also performed for other observed resonances. However, the correlations were considerably less well determined due most likely to the overlaps between near-by states or the very low statistics.

\subsection{Characteristic decay patterns}

\begin{figure*}
    \centering
    \subfigure{\label{fig:Decay1}}
    \vspace{-0.0in}
    \hspace{-0.0in}
    \subfigure{\label{fig:Decay2}}
    \includegraphics[width=.90\textwidth]{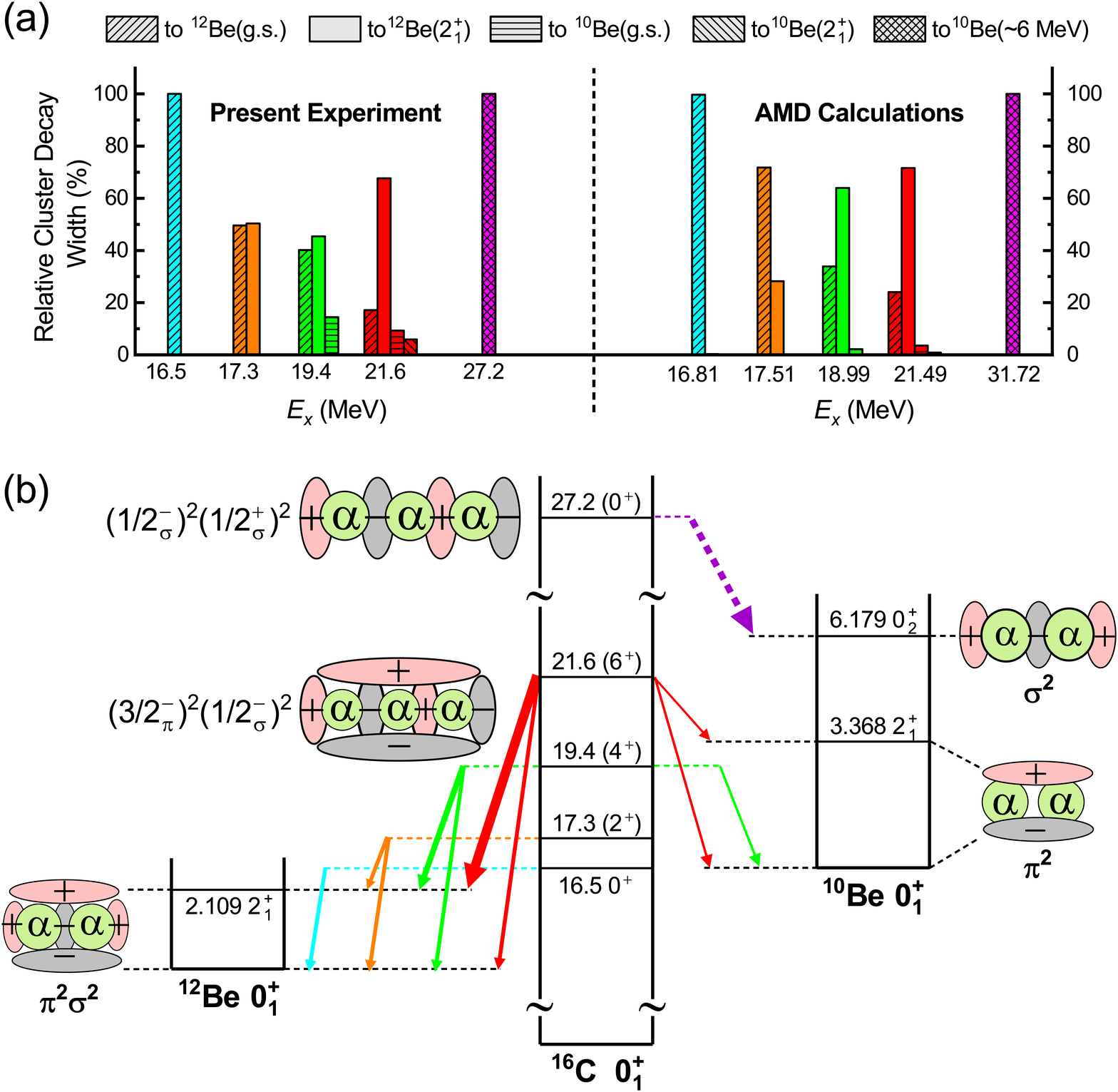}
    \caption{(a) Relative $\alpha$ and $^6$He decay widths from the resonances in $^{16}$C to the different final states of $^{12}$Be and $^{10}$Be, extracted from the presently observed spectra [Fig. \ref{fig:IM}] (left panel) and AMD calculations \cite{Baba2018,Baba2020} (right panel) \cite{Liu2020}. The sum of those widths from the same mother resonance is normalized to 100\%, and the various final states are distinguished by different colors. (b) Characteristic $\alpha$ and $^6$He decay pattern for the presently observed members of the positive-parity linear-chain molecular bands with $(3/2_\pi^-)^2(1/2_\sigma^-)^2$ and $(1/2^-_\sigma)^2(1/2_\sigma^+)^2$ configurations in $^{16}$C. For observed resonances in $^{16}$C between 16.5 and 21.6 MeV, the width of each arrow represents the cluster decay branching ratio relative to that decaying to the ground state of $^{12}$Be, which is also consistent with the AMD calculations \cite{Baba2018}. The width of the dotted arrow for higher resonance 27.2 MeV of $^{16}$C is just qualitative. The valence neutron configuration, symbolized by $\pi$- or $\sigma$-bond, for each chain-like structure is depicted according to the Ref. \cite{Baba2018,Baba2020}. The same decay paths presented in (a) and (b) are depicted in the same color. }
    \label{fig:Decay}
\end{figure*}

As the predicted feature of the linear-chain formation, we focus on the decay patterns of the resonances in $^{16}$C. 
From the reconstructed excitation energy spectra in Fig. \ref{fig:IM}, we extracted the relative cluster decay width for each resonant state in $^{16}$C, which is proportional to the number of counts in each fitted peak divided by the corresponding detection efficiency. The numerical results of the experiment and AMD calculations \cite{Baba2018,Baba2020} are given in Fig. \ref{fig:Decay1} which is adapted from Fig. 4 in our previous report \cite{Liu2020} and updated with the AMD calculated bandhead energy (31.72 MeV) of the pure $\sigma$-bond band according to Ref. \cite{Baba2020}. A schematic diagram is presented in Fig. \ref{fig:Decay2} to explicitly illustrate the observed decay patterns of $^{16}$C. 

The structural similarity between the mother and daughter states enhances dramatically decay process, as indicated in some earlier works \cite{Oertzen2006,Funaki2015} and predicted by the AMD calculations \cite{Baba2014,Baba2018,Baba2020}. Since clear molecular configurations have already been established in $^{10}$Be and $^{12}$Be \cite{Horiuchi2012,Freer2018,LiuYang2018,Liu2020}, the connected decay paths can be a strong signature of the similar bond structures in the mother nucleus $^{16}$C. 
For $^{12}$Be, the $(3/2_\pi^-)^2(1/2_\sigma^+)^2$ configuration is well established in the ground band states ($0^+_1$, $2^+_1$).
For $^{10}$Be, $\pi^2$ configuration is well developed in the ground band states ($0^+_1$, $2^+_1$), while $\sigma^2$ configuration is dominant in the $0^+_2$ state at around 6 MeV.
In the case of $^{16}$C, the AMD calculations \cite{Baba2018} predict a positive-parity linear-chain band with $(3/2_\pi^-)^2(1/2_\sigma^-)^2$ configuration for the four valence neutrons. The proposed members are at 16.81 ($0^+_6$), 17.51 ($2_9^+$), 18.99 ($4^+_{10}$), and 21.49 ($6^+_5$) MeV, as listed in Table. \ref{table:Levels} and displayed in the right panel of Fig. \ref{fig:Decay1}. These states should have large overlaps with the ground bands of both $^{10}$Be and $^{12}$Be, as illustrated in Fig. \ref{fig:Decay2}.
Decaying from this band into the higher excited band states of $^{10}$Be and $^{12}$Be is energetically prohibited. 
What is interesting here is the predicted increasing decay widths to the first excited state ($2_1^+$) in comparison to those to the ground state of $^{12}$Be,  as a function of the excitation energies and spins of the band members (from 16.81 MeV to 21.49 MeV) in $^{16}$C. The 21.49 MeV state in $^{16}$C was also predicted to decay to both $^{10}$Be($0^+_1$) and $^{10}$Be($2^+_1$) states with similar decay widths. These can be explained by the closer angular momentum (lower centrifugal potential) between the $2_1^+$ state in the daughter nucleus $^{12}$Be and the higher spin states in the mother nucleus $^{16}$C, in addition to the effect of decay energy, Coulomb barrier and structural link \cite{Baba2018,Baba2020}.

The presently observed states at 16.5, 17.3, 19.4 and 21.6 MeV are close in energy to the AMD predicted $0^+$, $2^+$, $4^+$ and $6^+$ members of the $(3/2_\pi^-)^2(1/2_\sigma^-)^2$-type band in $^{16}$C. The relative cluster decay widths extracted from the experiment for these states are also in excellent agreement with the predictions, as shown in Fig. \ref{fig:Decay1} and \ref{fig:Decay2} including the characteristic features as outlined above.
Most notably, the 21.6 MeV member decays much stronger to the $^{12}$Be($2^+_1$) state than to the $^{12}$Be($0^+_1$) state. 
Based on the above angular correlation analysis for the 16.5 MeV state and the observed decay pattern for other states, we have assigned the spin-parities of $0^+$, $2^+$, $4^+$ and $6^+$ for the 16.5, 17.3, 19.4 and 21.6 MeV resonances in $^{16}$C, respectively.  These resonances correspond to the four members of the AMD predicted positive-parity linear-chain molecular band with the $(3/2_\pi^-)^2(1/2_\sigma^-)^2$ configuration.

Another intriguing high-lying state at 27.2 MeV was observed to decay almost exclusively into $^{10}$Be($\sim$6 MeV), while its decay to the lower lying $^{10}$Be(g.s.) and $^{10}$Be($2^+_1$) states are negligible, as shown in Fig. \ref{fig:IMe} and Fig. \ref{fig:Decay}(a)(b). 
This unique decay path has recently been investigated by further AMD calculations \cite{Baba2020}, and a pure $\sigma$-bond linear-chain molecular band with $(1/2^-_\sigma)^2(1/2_\sigma^+)^2$ configuration has been predicted to appear at even higher energies in $^{16}$C. In contrast to the $(3/2_\pi^-)^2(1/2_\sigma^-)^2$ linear chain, the members of this band have been predicted to decay predominantly into the excited band states of $^{10}$Be($0^+_2$) and $^{12}$Be($0^+$, 13.6 MeV), but not to the ground band states.
This clearly distinguishes these two linear-chains. The decay pattern of the $\sigma$-bond linear chain can also be qualitatively understood from the structural link between the states in $^{10}$Be, $^{12}$Be and the linear-chain states in $^{16}$C \cite{Baba2020}.
Further experimental investigations are now planned in order to clarify the existence of this very high-lying $\sigma$-bond linear-chain molecular band in $^{16}$C.

\subsection{Energy-Spin systematics}

\begin{figure}
    \centering
    \includegraphics[width=.45\textwidth]{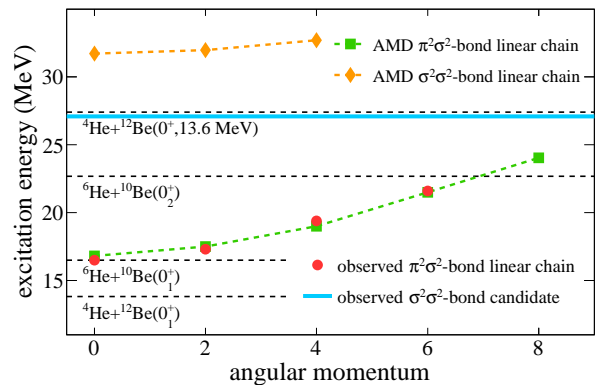}
    \caption{The energy-spin systematics of the molecular rotational bands. Red circles show the resonances observed in the present experiment and compared with the AMD predicted  $(3/2_\pi^-)^2(1/2_\sigma^-)^2$-bond linear-chain molecular band (green squares). The blue line shows the energy location of the observed state (27.2 MeV) which has a decay pattern similar to that of the predicted pure $\sigma$-bond linear-chain molecular band (Orange diamonds). }
    \label{fig:Sys}
\end{figure}

The excitation energies of the resonant states in $^{16}$C are plotted against spin $J$ in Fig. \ref{fig:Sys}, including the AMD predictions \cite{Baba2018,Baba2020} and present observations. Our observation of the 16.5 MeV ($0^+$), 17.3 MeV ($2^+$), 19.4 MeV ($4^+$) and 21.6 MeV ($6^+$) resonant states are in excellent agreement with the AMD predicted $\pi^2\sigma^2$-bond linear-chain molecular band \cite{Baba2018}.
From the experimental data, a ${\hbar^2}/{2\mathfrak{I}}$ value of about 122 keV can be deduced,  with $\mathfrak{I}$ being the moment of inertia. This value is comparable to the AMD calculated one (${\hbar^2}/{2\mathfrak{I}}=112$ keV) for the linear-chain molecular band and considerably smaller than those of the ground band (${\hbar^2}/{2\mathfrak{I}}=196$ keV) and the triangular-structure band  (${\hbar^2}/{2\mathfrak{I}}=238$ keV) \cite{Baba2018}. Newly calculated pure $\sigma$-bond linear-chain molecular rotational band is also plotted in the figure. The ${\hbar^2}/{2\mathfrak{I}}$ value of this band is estimated to be as small as about 50 keV, corresponding to an extremely large quadrupole deformation parameter $\beta$ of about 1.6 as the valence neutrons are mostly located between the cluster-cores \cite{Oertzen2006,Baba2020}. 

\section{SUMMARY}
In summary, we have carried out a new inelastic excitation and cluster-decay experiment using a $^{16}$C beam at about 23.5 MeV/nucleon off a $(\mathrm{CD}_2)_n$ target to investigate the linear-chain clustering structures in neutron-rich $^{16}$C. 
The triple coincident detection of the two decay fragments, $\rm{^4He + ^{12}Be}$ or $\rm{^6He+^{10}Be}$, together with the recoil particles $\rm{^2H}$ was realized with high detection efficiency, thanks to the specially designed detection setup including one multi-layer silicon telescope at around zero degrees. Although a large energy spread exists for the PF-type unstable nucleus beams, which often prohibits the measurement from obtaining a high resolution $Q$-value spectrum, the present triple coincident detection allows to deduce precisely the beam energy event-by-event. As a result, good $Q$-value resolutions were achieved for both $^{4}$He + $^{12}$Be and $^{6}$He + $^{10}$Be final decay channels and the $^{16}$C resonances can be reconstructed according to various decay paths. The observed resonances at 16.5(1), 17.3(2), 19.4(1) and 21.6(2) MeV were assigned as the $0^+$, $2^+$, $4^+$ and $6^+$ members, respectively, of the positive-parity $(3/2_\pi^-)^2(1/2_\sigma^-)^2$-bond linear-chain molecular band, based on the angular correlation analysis for the 16.5 MeV state and the excellent agreement of the decay patterns between the measurements and theoretical predictions. Moreover, another intriguing high-lying state was observed at 27.2 MeV which decays almost exclusively to the $\sim$ 6 MeV states of $^{10}$Be. This has stimulated further AMD calculation for the pure $\sigma$-bond linear-chain configuration in $^{16}$C. It would be very interesting to design new experiments to study resonances in $^{16}$C at even higher excitation domain in order to find this extremely long linear-chain structure. Moreover, it would be of great importance to clarify the negative-parity molecular bands in carbon isotopes.

\begin{acknowledgments}
The authors wish to thank the staffs of HIRFL-RIBLL for their technical and operational support. The discussion with Profs. Z. Z. Ren and F. R. Xu are gratefully acknowledged. This work has been supported by the National Key R\&D Program of China (Grant No. 2018YFA0404403) and the National Natural Science Foundation of China (Grants No. 11875074, No. 11875073,  No. 12027809, No. 11961141003, No. 11775004, and No. 11775003).
\end{acknowledgments}

\bibliography{References.bib}

\end{document}